\documentclass[12pt]{article}

\begin{document}

\title{Reinventing spacetime on a dynamical hypersurface}
\author{J. Ponce de Leon\thanks{E-mail: jponce@upracd.upr.clu.edu, jpdel1@hotmail.com}\\ Laboratory of Theoretical Physics, Department of Physics\\ 
University of Puerto Rico, P.O. Box 23343, San Juan, \\ PR 00931, USA} 
\date{March  2006}

\maketitle

\begin{abstract}
In braneworld models, Space-Time-Matter and other Kaluza-Klein theories,  
our spacetime is devised as a four-dimensional hypersurface {\it orthogonal} to the extra dimension    
in a five-dimensional bulk. 
We show that the FRW line element can be ``reinvented" on a dynamical four-dimensional hypersurface, which is {\it not} orthogonal to the extra dimension, without any internal contradiction. This hypersurface is selected by the requirement of continuity of the metric and depends explicitly on the evolution of the extra dimension. The main difference between the ``conventional" FRW,  on an orthogonal hypersurface,  and the new one is that the later contains higher-dimensional modifications to the regular  matter density and pressure in $4D$. We compare the evolution of the spacetime in  these two interpretations. We find that a wealth of ``new" physics can be derived  from a five-dimensional metric if it is  interpreted on a dynamical (non-orthogonal) $4D$ hypersurface. In particular, 
in the context of a well-known cosmological metric in $5D$, we construct a FRW model which is consistent with the late accelerated expansion of the universe, while fitting simultaneously the observational data for the deceleration parameter. The model predicts an effective equation of state for the universe, which is consistent with observations.

\end{abstract}

\medskip



\newpage

\section{Introduction}
Recently, there has been an increased interest in models where our four-dimensional universe is embedded in a higher-dimensional bulk spacetime having {\it large} extra dimensions. The scenario in these models is that matter fields are confined to our four-dimensional universe,  in a $1 + 3 + d$ dimensional spacetime, while gravity propagates in the extra $d$ dimensions as well \cite{Arkani1}-\cite{Arkani3}.

In gravitation and cosmology, much work has been done for $d = 1$.  In particular, braneworld models as well as the space-time-matter (STM) theory have become quite popular. 
In brane models the cylinder condition of the old Kaluza-Klein theory is replaced by the conjecture that the  ordinary matter and fields are confined to a four-dimensional subspace usually referred to as ``3-brane" \cite{RS2}-\cite{Maartens2}, which is a domain wall in a five-dimensional anti-de Sitter spacetime. In STM the conjecture is that the ordinary matter and fields that we observe in $4D$ result from the geometry of the extra dimension \cite{Wesson 1}-\cite{EMT}. This conjecture is supported by the fact that, as a consequence of Campbell's 
theorem \cite{Campbell}-\cite{Seahra},  solutions of the five-dimensional gravitational field
equations in apparent vacuum can always be interpreted as solutions of the 
four-dimensional field equations with matter. 

Although brane theory and STM have different physical motivations for the introduction of a large extra dimension, they share the same working scenario. Namely, (i) they allow the bulk metrics to have non-trivial dependence of the extra dimension. 
The general bulk metric ansatz is\footnote{Many authors choose $\Phi$ constant  and assume that the extra coordinate is spacelike, so that $\epsilon = - 1$.} 
\begin{equation}
\label{usual ansatz}
d{\cal S}^2 = g_{\mu\nu}(x^{\rho}, y)dx^{\mu}dx^{\nu} + \epsilon \Phi^2(x^{\rho}, y)dy^2.
\end{equation}
(ii) Our four-dimensional spacetime is identified  with a  hypersurface $\Sigma_{0}$, which is 
orthogonal 
to the extra dimension (in the sense that it's normal vector $n^A =  \delta^{A}_{4}/\Phi$ is tangent to the $y$-lines), and is 
located at some value of $y$ \cite{Shiromizu}-\cite{Jpdegr-qc/0111011}, e.g.,
\begin{equation}
\label{fixed brane}
y = y_{0}.
\end{equation}
(iii) The equations for gravity in $4D$ are the Einstein equations with an effective energy-momentum tensor\footnote{The effective matter quantities in $4D$ do not depend on whether we calculate them using the STM or the braneworld paradigm; only the interpretation is different in both theories.}. (iv) Matter fields and observers are unable to access the bulk. 

In this work we are concerned with point (ii) mentioned above. The main questions we ask here are: Why should we identify our  four-dimensional  spacetime with an  orthogonal hypersurface (\ref{fixed brane})? Why not to consider more general $4D$ hypersurfaces?. After all (\ref{fixed brane}) is an external condition  and {\it not} a requirement from the field equations. What is the physical criterion to decide which $4D$ hypersurface can, or cannot,  be identified with our spacetime? What would happen if we relax (\ref{fixed brane}) and consider that our universe can be generated  on a non-orthogonal (to $y$-lines) $4D$ hypersurface?. 

In this paper we discuss these questions in the context of homogeneous cosmologies in $5D$. As a  master criterion for identifying our physical spacetime we require the fulfillment of  the junction conditions. We show that these conditions are satisfied, not only on a  hypersurface $y = y_{0}$,  but also on a time-varying $4D$ hypersurface whose dynamics depends on the evolution of the extra dimension.

Before continuing, in order to avoid misunderstanding,  let us notice that in the  approach discussed in Refs. \cite{Ida}-\cite{Dadhich} our spacetime is described as a domain wall moving in a {\it static} five-dimensional bulk, which is the  $5D$ analog of the static Schwarzschild-anti-de Sitter spacetime. The consistency between that approach and the working scenario mentioned above is provided by the fact that there exists a coordinate transformation that brings the $5D$ line element of the static $Sch-AdS$ bulk used in \cite{Ida}-\cite{Dadhich} (with $\Phi \neq 1$, but $\dot{\Phi} = 0$) into the  bulk in gaussian normal coordinates used in \cite{Binetruy}-\cite{Vollick} (with $\Phi = 1$ and $\dot{\Phi} = 0$). Therefore both approaches are equivalent but in different systems of coordinates \cite{mukoyama2}.

We show that the effective, or total, matter induced on a time-varying\footnote{In what follows we use the short expressions ``fixed" and ``moving" hypersurface in order to refer to the orthogonal and non-orthogonal (to $y$-lines) hypersurfaces $\Sigma_{0}$ and $\Sigma$, respectively. } $4D$ hypersurface contains higher-dimensional modifications, due to the evolution of the extra dimension, which are absent  on a ``fixed" hypersurface (\ref{fixed brane}). We argue that the interpretation of $5D$ metrics on $y = y_{0}$  hypersurfaces eliminates the effects of the extra dimensions on $4D$, leading to minor, if any, departures from the dynamics predicted in four-dimensional general relativity.

Our aim is to compare and contrast the evolution of the conventional FRW spacetime, devised on a fixed $4D$ hypersurface,  with the evolution of the  FRW spacetime ``reinvented" on the moving hypersurface selected by the junction conditions. 

To this end we use a well-known cosmological metric in $5D$, which on $y = y_{0}$ reproduces the dynamics of the FRW models of general relativity. The spacetime model constructed on a  moving hypersurface presents a much richer dynamics;  there are two different scenarios. One of them can be used to represent an early universe, while the second scenario describes the recent  evolution of the universe. We find  that this later scenario (i) agrees with the observed accelerating universe, while fitting simultaneously the observational data for the deceleration parameter; and (ii) predicts an effective equation of state for the universe, which is consistent with observations. 

Our conclusion is that a wealth of ``new" physics can be derived  from a five-dimensional metric if it is  interpreted on a dynamical $4D$ hypersurface, as we suggest here.

\section{Homogeneous cosmology in $5D$}
In this scenario our homogeneous and isotropic universe is envisioned as embedded in a five-dimensional manifold with metric\footnote{Except for the factor $\Phi \neq 1$, this line element is widely used in brane and STM models to describe  homogeneous and spatially isotropic cosmologies in $5D$. See for example \cite{Maartens1}, \cite{Liu Wesson}, \cite{Jpdegr-qc/0111011}, \cite{Liu Mashhoon}.} 
\begin{equation}
\label{cosmological metric}
d{\cal{S}}^2 = N^2(t,y)dt^2 - A^2(t,y)\left[\frac{dr^2}{(1 - kr^2)} + r^2(d\theta^2 + \sin^2\theta d\phi^2)\right] + \epsilon \Phi^2(t, y)dy^2,
\end{equation}
where $k = 0, +1, -1$ and $t, r, \theta$ and $\phi$ are the usual coordinates for a spacetime with spherically symmetric spatial sections. We adopt signature $(+ - - - )$ for spacetime and the factor $\epsilon $ can be $- 1$ or $+ 1$ depending on whether the extra dimension is spacelike or timelike, respectively.

The usual approach is that FRW models  are recovered on some four-dimensional hypersurface $\Sigma_{0}: y = y_{0}$, orthogonal to the $y$-lines. The $4D$ metric on this hypersurface is then
\begin{equation}
\label{4D FRW models for any k}
ds^2_{|FRW} = dt^2 - a^2(t)\left[ \frac{dr^2}{1 - kr^2} + r^2 d\Omega^2\right].
\end{equation}
The junction conditions  require the metric to be continuous across $\Sigma_{0}$. Therefore,
\begin{equation}
\label{usual interpretation}
N(t, y_{0}) = N_{|\Sigma_{0}} = 1, \;\;\; A(t, y_{0}) = A_{|\Sigma_{0}} = a(t).
\end{equation}
With this interpretation the evolution  of the extra dimension, which is expressed through $\Phi$,  enters nowhere in the dynamics of the four-dimensional spacetime. In particular the effective matter quantities in $4D$ are exactly the same as in general relativity; it doesn't matter whether we calculate them using the STM or braneworld paradigm; only the interpretation is different in both theories.  

Let us now assume that our universe is generated on a ``moving" hypersurface 
\begin{equation}
\label{defining 4D}
\Sigma: y = f(t),
\end{equation}
which is not orthogonal to the extra dimension, in the sense that it's normal vector. 
\begin{equation}
n_{A} = \frac{\epsilon \Phi}{\sqrt{1 + \epsilon (\Phi \dot{f}/N)^2}}(- \dot{f}, 0, 0, 0, 1),
\end{equation}
is not tangent to the $y$-lines. 

The metric induced on $\Sigma$ is 
\begin{equation}
ds^2_{\Sigma} = \left[N^2(t, f(t)) + \epsilon \Phi^2 {\dot{f}}^2(t)\right]dt^2 - A^2(t, f(t))\left[ \frac{dr^2}{1 - kr^2} + r^2 d\Omega^2\right].
\end{equation}
By assumption, on  $\Sigma$ we should recover the  FRW line element (\ref{4D FRW models for any k}). The continuity of the $5D$ metric across $\Sigma$ demands
\begin{equation}
\label{equation}
\label{boundary conditions for moving Sigma}
\left[N^2(t, f(t)) + \epsilon \Phi^2 {\dot{f}}^2(t)\right] = 1,\;\;\;A^2(t, f(t)) = a^2(t).
\end{equation}
The first equation allows us to determine the hypersurface $f(t)$. This is a differential equation with two solutions. One of them  is the ``trivial" solution $f = f_{0} = constant$, which gives rise to the conventional FRW models given by (\ref{usual interpretation}). 

The second solution is the one important to us; it yields a different FRW universe where the scale factor depends on $f$, which in turn depends on $\Phi$.  
Therefore, with this interpretation, the dynamics  in $4D$ crucially depends on the evolution of the extra dimension. Thus, it carries higher dimensional modifications to general relativity. 

These modifications become evident when we study the total, or effective, induced matter quantities\footnote{As it was mentioned in the Introduction, in point (iii), the effective matter quantities are obtained from the Einstein equations in $4D$.}. Specifically, we find that the effective matter density $\rho_{eff}$ can be separated in two parts
\begin{equation}
\rho_{eff} = \rho_{0} + \rho_{f}.
\end{equation}
where
\begin{eqnarray}
\label{matter density}
8 \pi G \rho_{0} &=& 3\left(\frac{\dot{a}}{a}\right)^2 + \frac{3 k}{a^2},\nonumber \\
8 \pi G \rho_{f} &=& 3\left[\frac{2 {\dot a}a'}{a^2}{\dot f} + \left(\frac{a'}{a}\right)^2 {\dot f}^2\right].
\end{eqnarray}
Here $\dot{a} = \partial A/\partial t$ and $a' = \partial A/\partial y$, both evaluated at $y = f(t)$, so matter quantities depend on $t$, but {\it not}\footnote{This is the requirement (iv) mentioned in the Introduction} in $y$.
The first term $\rho_{0}$ is the usual general relativity term recovered on $y = y_{0}$ hypersurfaces. The second one is the  higher dimensional correction, which depends on the  evolution of the extra dimension, viz.,
\begin{equation}
\label{explicit dependence on the extra dim}
\dot{f} = \sqrt{\frac{N^2 - 1}{\epsilon \Phi^2}}.
\end{equation} 
Similarly, the effective pressure can be separated as
\begin{equation}
p_{eff} = p_{0} + p_{f},
\end{equation}
with
\begin{eqnarray}
\label{matter pressure}
8 \pi G p_{0} = - \left(\frac{2 \ddot{a}}{a} + \frac{{\dot{a}}^2}{a^2}\right) - \frac{k}{a^2},
\end{eqnarray}
\begin{equation}
8 \pi G p_{f} = - \left[\left( \frac{4 {\dot{a}}'}{a}  
+ \frac{ 2\dot{a}a'}{a^2}\right)\dot{f} 
+ \left(\frac{2 a''}{a} + \frac{{a'}^2}{a^2}\right) {\dot{f}}^2
+ \left(\frac{2a'}{a}\right)\ddot{f}\right].
\end{equation}
Thus, if we take $f = constant$, then the effective matter content  is the same as in four-dimensional general relativity. Therefore, we conclude that the choice $y = y_{0}$ suppresses the effects of the extra dimension on the dynamics of $4D$.  There are only minor departures in the interpretation of the effective matter quantities, depending on the theory, but this does not  affect  the evolution of the universe, which depends on the total matter and pressure.

If $f \neq constant$, then $\rho_{f}$ and $p_{f}$ play a crucial role in the evolution of our $4D$ universe. 

\section{The standard cosmological model}

In order to  get a feeling of the ``new" physics associated with the higher dimensional modifications carried by the choice $f \neq constant$, we will consider the well-known $5D$ metric \cite{JPdeL 1}
\begin{equation}
\label{Ponce de Leon solution}
d{\cal S}^2 = y^2 dt^2 - A^2 t^{2/\alpha}y^{2/(1 - \alpha)}[dr^2 + r^2(d\theta^2 + \sin^2\theta d\phi^2)] - \alpha^2(1- \alpha)^{-2} t^2 dy^2,
\end{equation}
where $\alpha$ is a dimensionless parameter, and $A$ is an arbitrary constant with dimension $L^{- 1/\alpha}$. The extra coordinate $y$ is taken to be dimensionless. This is a solution of the five-dimensional Einstein equations in $5D$ vacuum, i.e., $^{(5)}T_{AB} = 0$. 

On the hypersurface $\Sigma: y = f(t)$, the induced $4D$ metric is
\begin{equation}
ds^2_{\Sigma} = \left(f^2 - \frac{\alpha^2}{(1 - \alpha)^2}t^2 \dot{f}^2 \right)dt^2 - A^2 t^{2/\alpha}y^{2/(1 - \alpha)}[dr^2 + r^2(d\theta^2 + \sin^2\theta d\phi^2)]
\end{equation}
The junction conditions  require the metric to be continuous across $\Sigma$. Specifically, on $\Sigma$
\begin{equation}
\label{FRW from the brane}
ds^2_{\Sigma} = dt^2 - a^2(t)\left[ dr^2 + r^2 d\Omega^2\right],
\end{equation}
for spatially flat FRW models. Thus, 
\begin{equation}
\label{defining f}
f^2 - \frac{\alpha^2}{(1 - \alpha)^2}t^2 \dot{f}^2 = 1,
\end{equation}
and 
\begin{equation}
a(t) = At^{1/\alpha}f^{1/(1 - \alpha)}.
\end{equation}
Equation (\ref{defining f}) has two solutions. The simplest one is $f = 1$, which corresponds to the usual interpretation. Namely, 
on the hypersurface $\Sigma_{0}: y = y_{0} = 1$, this metric corresponds to Friedmann-Robertson-Walker models with flat $3D$ sections. For the case under consideration, using (\ref{Ponce de Leon solution}) (\ref{matter density}) and (\ref{matter pressure}) we obtain
\begin{equation}
\label{density on y = const}
8 \pi G \rho_{0} = \frac{3}{\alpha^2 t^2},\;\;\;\mbox{and}\;\;\;8 \pi G p_{0} = \frac{(2\alpha - 3)}{\alpha^2 t^2}.
\end{equation}
The equation of state of the effective perfect fluid in $4D$ is: $p = \gamma \rho$ with $\gamma = (2\alpha/3 -1)$ ($\alpha = 2$ for radiation, $\alpha = 3/2$ for dust, etc.).

 The other solution to (\ref{defining f}) is
\begin{equation}
f(t) = \frac{1}{2}\left\{C t^{(1 - \alpha)/\alpha} + \frac{1}{C t^{(1 - \alpha)/\alpha}}\right\},
\end{equation}
where $C$ is a constant of integration with units of $L^{(\alpha - 1)/\alpha}$. We note that we cannot set $C = 0$. So, there is no continuous connection between this and the $y = 1$ solution.

A simple inspection shows that the behavior of the cosmic scale factor crucially depends on whether $\alpha > 1$ or $\alpha < 1$. We now proceed to investigate them separately.

\subsection{The solution for $ \alpha > 1$ } 
The cosmic scale factor is given by
\begin{equation}
\label{a1}
a_{1}(t) = A_{1}(2C_{1})^{1/(\alpha - 1)} \frac{t^{2/\alpha}}{\left[C_{1}^2 + t^{2(\alpha - 1)/\alpha}\right]^{1/(\alpha - 1)}}. 
\end{equation}
In the very early universe, for small values of $t$,  we have $a_{1} \approx t^{2/\alpha}$, and the corresponding matter distribution is
\begin{equation}
\label{matter quantities for the very early universe}
8\pi G\rho_{eff} = \frac{12}{\alpha^2 t^2}, \;\;\;8\pi Gp_{eff} = \frac{4(\alpha - 3)}{\alpha^2t^2},\;\;\;\frac{p_{eff}}{\rho_{eff}} = \frac{\alpha - 3}{3}.
\end{equation} 
The gravitational density of the induced matter is 
\begin{equation}
\label{gravitationa density for primordial matter}
(\rho_{eff} + 3p_{eff}) = (\alpha - 2)\rho_{eff}, 
\end{equation}
and the deceleration parameter $q$ is
\begin{equation}
\label{q asymptotically}
q = - \frac{\ddot{a}a}{{\dot{a}}^2} = \frac{1}{2}(\alpha - 2).
\end{equation}
The dominant energy condition applied to primordial matter confines the possible values of  $\alpha$. Namely,  
\begin{equation}
\label{range of alpha}
1 < \alpha \leq 6.
\end{equation}
The above shows that (i) for $\alpha > 2$ the effective  primordial matter behaves similar to ordinary gravitating matter (like radiation  for $\alpha = 4$, or stiff matter for $\alpha = 6$) where  the expansion is slowing down; (ii) for $\alpha = 2$ the effective primordial matter behaves like a network of cosmic strings $(p_{eff} = - \rho_{eff}/3)$ and (iii)   
for $1 < \alpha < 2$  the primordial cosmological ``fluid" has repulsive properties; it violates the strong energy condition $(\rho_{eff} + 3p_{eff}) > 0$ and the deceleration parameter is negative. 

A remarkable feature of this solution is that for large values of $t$, $a_{1}$ tends to a constant value, viz.,  $a_{1}(t) \rightarrow A_{1}(2C_{1})^{1/(\alpha - 1)}$, asymptotically. In terms of the redshift $z$;  the coordinate $t$ used in $a_{1}$ does not cover the whole range $z \in (0, \infty)$; where $z = 0$ and $z = \infty$ represent the redshift today and at the big bang, respectively.  Therefore, (\ref{a1}) describes the evolution of the universe from $z = \infty$
to some $\stackrel{\ast}{z} > 0$, which depends on $\alpha$.
\subsection{The solution for $\beta =  \alpha < 1$ }

In what follows, in order to avoid any confusion we will use $\beta$ to denote $\alpha < 1$. With this notation the cosmic scale factor is given by
\begin{equation}
\label{a2}
a_{2}(t) = A_{2}(1/2C_{2})^{1/(1 - \beta)}\left[1 + C_{2}^2t^{2(1 - \beta)/\beta} \right]^{1/(1 - \beta)}.
\end{equation}
We observe that the constants here are different from those in (\ref{a1}). In particular they have different units. Namely, 
\begin{equation}
\label{units}
C_{1} \sim L^{(\alpha - 1)/\alpha}, \;\;\;A_{1} \sim L^{- 1/\alpha},\;\;\; C_{2} \sim L^{-(1 - \beta)/\beta},\;\;\;  A_{2} \sim L^{- 1/\beta}.
\end{equation}
For $t = 0$, $a_{2}(t) = A_{2}(1/2C_{2})^{1/(1 - \beta)}$. So, if we choose the constants in such a way that  $A_{1}(2C_{1})^{1/(\alpha - 1)} = A_{2}(1/2C_{2})^{1/(1 - \beta)}$, then $a_{2}$ starts exactly where $a_{1}$ ends. The conclusion is that $a_{1}$ and $a_{2}$, represent {\it different} stages of the evolution of the universe; $a_{1}$ can be used to represent an early evolution from $z = \infty$ to $z = \stackrel{\ast}{z} > 0$, while $a_{2}$ corresponds to late stages of the  evolution from $z = \stackrel{\ast}{z} > 0$ to $z = 0$. 

In principle, one can obtain  $\stackrel{\ast}{z}$ in terms of $\alpha$ and $\beta$ by joining the metrics $a_{1}$ and $a_{2}$ across a $t = \stackrel{\ast}{t} = constant$ hypersurface. However, such calculation is beyond the scope of this paper.

For this solution the deceleration parameter is 
\begin{equation}
\label{q2}
q_{2} = \frac{1}{2}(\beta - 2) + \frac{(3\beta -2)}{2 y},
\end{equation}
where we have used the dimensionless coordinate $y$
\begin{equation}
\label{y}
y = C_{2}^2 t^{2{(1 - \beta)/\beta}}.
\end{equation}
Clearly for ``large" values of $t$, $q_{2}$ becomes negative, meaning that the expansion is speeding up. Evidence in favor of a recent accelerated expansion is provided by  observations of high-redshift supernovae Ia \cite{Riess}-\cite{Tonry}, as well as  other observations, including  the cosmic microwave background and galaxy power  spectra \cite{Lee}-\cite{Sievers}.

Let us study this in more detail. The effective matter quantities are
\begin{eqnarray}
\label{effective quantities}
 8 \pi G \rho_{eff} &=& \frac{12y^2}{\beta^2 t^2(1 + y)^2},\nonumber \\
8\pi G p_{eff} &=& \frac{4(\beta - 3)y^2 + 4(3\beta - 2)y}{\beta^2 t^2(1 + y)^2}. 
\end{eqnarray}
For large values of $t$, or $y$, the effective pressure is negative. In order to interpret this, the usual approach is to invoke the existence of some kind of matter, sometimes  called missing  or dark energy, which possesses a large negative pressure \cite{Pelmutter2}. The simplest candidate for this  energy is the cosmological constant \cite{Riess}, \cite{Peebles}, \cite{Padmanabhan0}, which looks like an ideal fluid with negative pressure  $p_{\Lambda} = - \rho_{\Lambda}$. 

Therefore, we assume that the universe is filled with ordinary matter and a cosmological term, i.e., 
\begin{eqnarray}
\label{splitting the effective quantities}
8\pi G\rho_{eff} &=& 8\pi G\rho_{m} + \Lambda,\nonumber \\ 
8\pi Gp_{eff} &=& 8\pi Gp_{m} - \Lambda.
\end{eqnarray}
Since the overwhelming time of the evolution of the universe is spent in the matter-dominated domain, we set $p_{m} = 0$ and  find
\begin{eqnarray}
\label{density for the regular matter}
8 \pi G \rho_{m} = \frac{4\beta y^2 + 4(3\beta - 2)y}{\beta^2 t^2(1 + y)^2},
\end{eqnarray}
and
\begin{equation}
\label{cosmological term}
\Lambda = 4 \frac{(3 - \beta)y^2 -(3\beta - 2)y}{\beta^2 t^2(1 + y)^2}.
\end{equation}
An important quantity in cosmology is the so called density parameter $\Omega_{m}$, which is  
\begin{equation}
\label{density parameter}
\Omega_{m} = \frac{8 \pi G \rho_{m}}{3H^2}= \frac{1}{3}\left[{\beta +(3\beta - 2)y^{- 1}}\right].
\end{equation}
This expression leads to interesting results.
\subsubsection{Relation between observables $q$ and $\Omega_{m}$:} Substituting (\ref{density parameter}) into $q_{2}$ from (\ref{q2}), we get
\begin{equation}
\label{general formula for the acceleration}
q_{2} = -1 + \frac{3}{2} \Omega_{m},
\end{equation}
regardless of $\beta$. In particular, if we take ${\bar{\Omega}}_{m} = 0.3$ today, then we obtain
\begin{equation}
\bar{q}_{2} = - 0.55,
\end{equation}
for the present acceleration, which is within the region of the suspected values for the present acceleration of the universe.

\subsubsection{Equation of state of the universe:} For the total, or effective,  energy density and pressure the equation of state of the universe can be written as
\begin{equation}
\label{definition of w}
w = \frac{p_{eff}}{\rho_{eff}}.
\end{equation}
Then, from (\ref{effective quantities}) and (\ref{density parameter}) we find
\begin{equation}
\label{equation of state for the universe}
w = -1 + \Omega_{m},
\end{equation}
which for ${\bar{\Omega}}_{m} = 0.3$ gives $w = - 0.7$ in agreement with observations. Indeed, this result is similar to the one derived for quintessence \cite{Efstathiou}. In this framework, observations from  SNe Ia and CMB indicate that the equation of state for quintessence $w_{Q} = p_{Q}/\rho_{Q}$ has an upper limit $w_{Q} \approx - 0.6$, which is close to the lower limit $w_{Q} \approx - 0.7$ allowed for  quintessence tracker fields \cite{Steinhard}. 

\subsection{Example: $\beta = 2/3$}
An important question is whether the effective matter satisfies some physical conditions.  We know that the strong energy condition has been violated in different stages of the evolution of the universe; first during inflation and now during the present accelerated expansion of the universe. However, we can still require (i) that the total energy density and pressure satisfy the dominant energy condition; (ii) positivity of the matter energy density; (iii) positivity of the cosmological term. We consider any kind of ``exotic" matter not satisfying these conditions as ``unphysical".

There are many choices of  $\beta$ for which the matter, during the epoch described by $a_{2}$, satisfies the above-mentioned  physical conditions. However, here we will focus on the simplest example,  which is for $\beta = 2/3$. 
In this case $y = C_{2}^{2}t$, the deceleration is constant $q = -2/3$ and $\Omega_{m} = 2/9 \approx 0.22$, which are in the region of allowed values. The total matter and pressure
\begin{equation}
\label{effective quantities for beta = 2/3}
8\pi G\rho_{eff} = \frac{27 C_{2}^4}{(1 + C_{2}^2 t)^2}, \;\;\;8\pi Gp_{eff} = - \frac{21 C_{2}^4}{(1 + C_{2}^2 t)^2},
\end{equation}
satisfy the dominant energy condition. The matter density and cosmological term are positive as required, viz.,
\begin{equation}
\label{effective quantities for beta = 2/3}
8\pi G\rho_{m} = \frac{6 C_{2}^4}{(1 + C_{2}^2 t)^2}, \;\;\;\Lambda =  \frac{21 C_{2}^4}{(1 + C_{2}^2 t)^2}.
\end{equation}
We would like to point out, the obvious fact, that the separation of effective quantities into components is not unique.  For example, we could  assume that the effective matter is the superposition of several fluids with distinct equations of state. What is important here is that only the effective (or total) quantities have observational consequences.

\section{Summary and discussion}
In this paper we have shown that the familiar FRW line element can be recovered not only on a orthogonal hypersurface $\Sigma_{0}: y = y_{0}$, but also on a non-orthogonal one $\Sigma : y = f(t)$, without any internal contradiction.

The ``physical mechanism" for the choice of the four-dimensional subspace is provided by the continuity of  the $5D$ metric across $\Sigma: y = f(t)$, which yields (\ref{boundary conditions for moving Sigma}).
Thus, the same mechanism that leads to the choice $f = f_{0} = constant$, is the mechanism that allows $f \neq constant$.  Usually,  the embedding  $f = f_{0} = constant$ is the only one considered in the literature. 
Therefore,  the motivation of this paper has been to study the feasibility and consequences of  the second possibility $f \neq constant$ given by (\ref{explicit dependence on the extra dim}). 

 The effective energy-momentum tensor in four-dimensions is defined through the Einstein field equations in $4D$, namely 
\begin{equation}
\label{effective matter}
^{(4)}G_{\mu\nu} = 8 \pi G T^{(eff)}_{\mu\nu},
\end{equation}
which is indicated in the Introduction as point (iii). This effective energy-momentum tensor is the one that governs the dynamics in $4D$. As a consequence of the contracted Bianchi identities in $4D$, ${^{(4)}G}^{\mu}_{\nu; \mu} = 0$, the effective energy-momentum tensor satisfies the standard\footnote{If in the bulk there were scalar and/or other fields,  other than ${^{(5)}T}_{AB} = \Lambda_{(5)}\gamma_{AB},
$  then this would be no longer true,  in general.
} general relativity conservation equations, viz., 
\begin{equation}
\nabla^{\mu}T_{\mu\nu}^{(eff)} = 0.
\end{equation}
Thus, for any solution of Einstein's field equations in $5D$ one can define an effective energy/matter density and pressure\footnote{Mathematically, it is now accepted that Campbell's theorem
guarantees that any solution of the four-dimensional Einstein field
equations with matter can be derived from a solution of the five-dimensional
Kaluga-Klein equations in apparent vacuum, provided an appropriate
identification is made for the energy-momentum tensor}. The effective equations in $4D$ (\ref{effective matter}) are the same in braneworld models as well as in STM \cite{Jpdegr-qc/0111011}, just the interpretation is different\footnote{It is well known that in GR, the same energy-momentum tensor can be interpreted in different ways. In brane models, the assumption is that our universe is a singular surface in an Anti-de Sitter universe of five dimensions. In this case Israel's boundary conditions are used to define an energy-momentum $T_{\mu\nu}^{Israel}$. Then $T_{\mu\nu}^{eff} = T_{\mu\nu}^{Israel} + $ quadratic ``corrections" + dark (or Weyl) energy (radiation). In STM, $T_{\mu\nu}^{eff}$ is usually interpreted as the superposition of fluids with different equations of state.}.  It should be emphasized that the interpretation of the effective energy-momentum {\it does not} affect the evolution of the universe.

Therefore, in the present paper we have adopted a ``generic" approach and studied the physics of the effective matter quantities on $\Sigma$, defined through the Einstein equations in $4D$. They  contain  not only the ``ordinary" $\rho_{0}$ and $p_{0}$ matter terms of general relativity, recovered on the orthogonal $\Sigma_{0}$ hypersurface, but also $\rho_{f}$ and $p_{f}$ which look as the density and pressure of some ``new" matter. This new matter is an effect from the gravitational field in the  bulk, transmitted to $4D$ through the dynamics of $\Sigma: y = f(t)$. In the conventional FRW interpretation, for which $y = y_{0}$, our spacetime is disconnected  from the extra dimension in the sense that these new terms do not appear.

Thus,  a wealth of new physics can be derived  from a five-dimensional metric if we interpret it on a dynamical (non-orthogonal) $4D$ hypersurface, as we suggest here. Our discussion in section $3$  clearly illustrates this point. Namely, the cosmological metric (\ref{Ponce de Leon solution}), when interpreted on a fixed hypersurface $\Sigma_{0}: y = y_{0}$, corresponds to familiar power-law flat-FRW models (\ref{density on y = const}) in $4D$. Nothing else; there are no dynamical effects coming from the extra dimension.

However, a very different physical picture is obtained when the FRW line element is recovered from (\ref{Ponce de Leon solution}) on $\Sigma : y = f(t)$. In this case, the evolution of the universe can be separated into three stages.
The first one is the early universe, which can be described by the solution $a_{1}(t)$, as discussed in section $3.1$. As the universe expands and ages, $a_{1}(t) \rightarrow A_{1}(2C_{1})^{1/(\alpha - 1)}$, asymptotically. Consequently, the second stage corresponds to the end of the era described by $a_{1}$ and the beginning of the era described by $a_{2}$.  In the  third stage, which is described by $a_{2}$,  the universe can be considered as a mixture of dust and a variable cosmological ``constant". In this epoch the universe is dominated by the vacuum energy,  which is responsible for the observed present acceleration.

It is important to note that similar results are obtained from very different approaches. For example,  the effective equation of state of the universe, for the total pressure and density. Our result (\ref{equation of state for the universe}),  is quintessence-like namely, $w \approx - 0.7$, for $\Omega_{m} \approx 0.3$, which coincides with the one obtained from other models \cite{Efstathiou}-\cite{Steinhard}. 

Also, if we use that $\Omega_{m} + \Omega_{\Lambda} = 1$, then our formulae for the deceleration parameter (\ref{general formula for the acceleration}) and the equation of state of the universe (\ref{equation of state for the universe}) can be written as 
\begin{equation}
q = 2 - \frac{3}{2}(\Omega_{m} + 2\Omega_{\Lambda}),
\end{equation}
and
\begin{equation}
w =  1 - \Omega_{m} - 2\Omega_{\Lambda}.
\end{equation}
These expressions are {\it identical} to those obtained in braneworld models with variable vacuum energy \cite{JPdeLgr-qc/0401026}, on a fixed hypersurface 
$\Sigma_{0}: y = y_{0}$. They reduce to the appropriate FRW-counterparts for $\Omega_{\Lambda} = 0$ and $\Omega_{m} = 1$. In particular we obtain $q = 1/2$ and $w = 0$ as in the dust-FRW cosmologies.

We would like to finish this paper with the following comment. The cosmological metrics given by $a_{1}$ and $a_{2}$ are interesting in their own right, regardless of the way they were obtained. A more detailed investigation of these cosmologies is presented in \cite{JPdeLgr-qc/0511150}, where  we  match them  across a $t = \stackrel{\ast}{t} = constant$ hypersurface and obtain, among other things, a detailed information of the ``second" stage of evolution mentioned above.


\begin{thebibliography}{99}
\bibitem{Arkani1}{N. Arkani-Hamed, S. Dimipoulos and G. Dvali, {\em Phys. Lett.} {\bf B 429}, 263(1998), hep-ph/9803315.}
\bibitem{Arkani2}{N. Arkani-Hamed, S. Dimipoulos and G. Dvali, {\em Phys. Rev.} {\bf D 59}, 086004(1999), hep-ph/9807344.}
\bibitem{Arkani3}{I. Antoniadis, N. Arkani-Hamed, S. Dimipoulos and G. Dvali, {\em Phys. Lett.} {\bf B 436}, 257(1998), hep-ph/9804398.}
\bibitem{RS2}{L. Randall and R. Sundrum, {\em Phys. Rev. Lett.} {\bf 83}, 4690(1999), hp-th/9906064.}
\bibitem{Maartens1}{Roy Maartens, ``Geometry and dynamics of the brane-world", Reference Frames and Gravitomagnetism, ed. J Pascual-Sanchez {\em et al}. (World Sci., 2001), p93-119, gr-qc/0101059.}
\bibitem{Maartens2}{R. Maartens, {\em Living Rev.Rel.} {\bf 7}, 7(2004), gr-qc/0312059.}


\bibitem{Wesson 1}{P.S. Wesson, {\em G. Rel. Gravit.} {\bf 16}, 193(1984).}
\bibitem{JPdeL 1}{J. Ponce de Leon, {\em Gen. Rel. Grav.} {\bf 20}, 539(1988).}
\bibitem{WessonJPdeL}{P.S. Wesson and J. Ponce de Leon, {\em J. Math. Phys.} {\bf 33}, 3883(1992).}
\bibitem{JPdeL Wesson}{J. Ponce de Leon and P.S. Wesson, {\em J. Math. Phys.} {\bf 34}, 4080(1993).}
\bibitem{Coley1}{A.A. Coley and D.J. McManus, {\em J. Math. Phys.} {\bf 36}, 335(1995).}

\bibitem{Wesson book}{P.S. Wesson, {\em Space-Time-Matter} (World Scientific Publishing Co. Pte. Ltd. 1999).}
\bibitem{EMT}{J. Ponce de Leon, {\em Int.J.Mod.Phys.} {\bf D11}, 1355(2002), gr-qc/0105120.}
\bibitem{Campbell}{J.E. Campbell, {\em A Course of Differential Geometry} 
(Clarendon, Oxford, 1926).}
\bibitem{Rippl}{S. Rippl, C. Romero and R. Tavakol, {\em Class. Quant. 
Grav.} {\bf 12}, 2411(1995), gr-qc/9511016.}
\bibitem{Romero}{C. Romero, R. Tavakol and R. Zalaletdinov, {\em Gen. Rel. 
Grav.} {\bf 28}, 365(1996).}
\bibitem{Lidsey}{J.E. Lidsey, C. Romero, R. Tavakol and S. Rippl, {\em 
Class. Quant. Grav.} {\bf 14}, 865(1997), gr-qc/9907040.}
\bibitem{Seahra}{S.S. Seahra and P.S. Wesson, {\em Class. Quant. Grav.} {\bf 
20}, 1321(2003), gr-qc/0302015.}

\bibitem{Shiromizu}{T. Shiromizu, K. Maeda and M. Sasaki, {\em Phys. Rev.} {\bf D62}, 02412(2000), gr-qc/9910076.}
\bibitem{Liu Wesson}{H. Liu and P.S. Wesson, {\em Astrophys. J.} {\bf 562}, 1(2001), gr-qc/0107093.}
\bibitem{Chang}{Baorong Chang, Hongya Liu, Huanying Liu, Lixin Xu, {\em Mod.Phys.Lett.} {\bf A20}, 923(2005), astro-ph/0405084.}
\bibitem{Jpdegr-qc/0111011}{J. Ponce de Leon, {\em Mod.Phys.Lett.} {\bf A16},  
2291(2001), gr-qc/0111011.}
\bibitem{Ida}{D. Ida, {\em JHEP} {\bf 0009}, 014(2000),  gr-qc/9912002.}
\bibitem{Kraus}{P. Kraus, {\em JHEP}  {\bf 9912}, 011(1999), hep-th/9910149.}
\bibitem{Kaloper}{N. Kaloper, {\em Phys.Rev} {\bf D60}, 123506(1999), hep-th/9905210.}
\bibitem{Barcelo}{C. Barcelo, M. Visser, {\em Nucl.Phys.}  {\bf B584}, 415(2000), hep-th/0004022.}
\bibitem{Dadhich}{P. Singh, N. Dadhich, {\em Mod.Phys.Lett.} {\bf A18}, 983(2003), hep-th/0204190; hep-th/0208080.}
\bibitem{Binetruy}{P. Binetruy, C. Deffayet, U. Ellwanger, D. Langlois, {
Phys.Lett.}  {\bf B477}, 285(2000), hep-th/9910219.}
\bibitem{mukoyama1}{S. Mukohyama, {\em Phys.Lett.} {\bf B473}, 241(2000), hep-th/9911165.}
\bibitem{Vollick}{D. N. Vollick, {\em  Class.Quant.Grav.} {\bf 18}, 1(2001), hep-th/9911181.}
\bibitem{mukoyama2}{S. Mukohyama, T. Shiromizu and K. Maeda, {\em Phys.Rev.},  {\bf D62} 024028(2000), hep-th/9912287. }
\bibitem{Liu Mashhoon}{H. Liu and B. Mashhoon, {\em Ann. Phys.} {bf 4}, 565(1995).}

\bibitem{Riess}{A.G. Riess {\it et al.,}  Supernova Search Team Collaboration, {\em Astron. J.}, {\bf 116}, 1009 (1998),  astro-ph/9805201.}
\bibitem{Perlmutter}{S. Perlmutter {\it et al.,} Supernova Cosmology Project Collaboration,   {\em Astrophys. J.}, {\bf 517},
565 (1999), astro-ph/9812133.}

\bibitem{Liddle}{Andrew R Liddle,  {\em New Astron.Rev.}, {\bf 45}, 235(2001), astro-ph/0009491.}
\bibitem{Seto}{N. Seto, S. Kawamura and T. Nakamura,  {\em Phys.Rev.Lett.} {\bf 87}, 221103(2001), astro-ph/0108011.}

\bibitem{Knop}{R. A. Knop {\em et al},  {\em Astrophys. J.}, {\bf 598},  102(2003), astro-ph/0309368.}
\bibitem{Tonry}{J.L. Tonry {\em et al}, {\em Astrophys. J.}, {\bf  594},  1 (2003), astro-ph/0305008.}

\bibitem{Lee}{A.T. Lee et al, {\em Astrophys. J.}, {\bf 561},  L1(2001), astro-ph/0104459.}


\bibitem{Stompor}{R. Stompor et al, {\em Astrophys. J.}, {\bf 561},  L7(2001), astro-ph/0105062.}
\bibitem{Halverson}{
N.W. Halverson et al,  {\em Astrophys. J.}, {\bf 568},  38(2002), astro-ph/0104489.}

\bibitem{Netterfielf}{C.B. Netterfield et al,   {\em Astrophys. J.}, {\bf 571}, 604(2002), astro-ph/0104460.}

\bibitem{Pryke}{C. Pryke, {\it et al.,}  {\em Astrophys. J.}, {\bf 568},  46(2002),  astro-ph/0104490.}
\bibitem{Spergel}{D.N. Spergel {\it et al.,} {\em Astrophys. J.Suppl.},  {\bf 148}, 175 (2003), astro-ph/0302209.}
\bibitem{Sievers}{J. L. Sievers, {\it et al.,}  {\em Astrophys. J.},  {\bf 591},  599(2003), astro-ph/0205387.}

\bibitem{Pelmutter2}{S. Perlmutter, M. S. Turner and M. White, {\em Phys.Rev.Lett.} {\bf 83}, 670(1999), astro-ph/9901052.} 

\bibitem{Peebles}{P. J. E. Peebles and  B. Ratra, {\em Rev.Mod.Phys.} {\bf 75}, 559(2003), astro-ph/0207347.}
\bibitem{Padmanabhan0}{T. Padmanabhan, {\em Phys.Rept.} {\bf 380},  235(2003), hep-th/0212290.}
\bibitem{Efstathiou}{G. Efstathiou, ``Constraining the equation of state of the Universe from Distant Type Ia Supernovae and Cosmic Microwave Background Anisotropies", astro-ph/9904356.}
\bibitem{Steinhard}{P.J. Steinhardt, L. Wang and I. Zlatev,  {\em Phys.Rev.} {\bf D59},  123504(1999),  astro-ph/9812313.}

\bibitem{JPdeLgr-qc/0401026}{J. Ponce de Leon, {\em Gen. Rel. Grav.} {\bf 
37}, 53(2005), gr-qc/0401026.}
\bibitem{JPdeLgr-qc/0511150}{J. Ponce de Leon, ``An analytic model for the transition from decelerated to accelerated cosmic expansion", gr-qc/0511150.}



\end{thebibliography}
\end{document}